\documentclass[preprint]{aastex61}
\usepackage{amsmath}
\usepackage{amssymb}
\usepackage{graphicx}

\submitjournal{ApJ}

\shorttitle{Multi-frequency scattering of low DM pulsars}
\shortauthors{Krishnakumar et al.}

\begin{document}

\title{Multi-frequency scatter broadening evolution of pulsars - II. Scatter broadening 
of nearby pulsars}

\correspondingauthor{M.A. Krishnakumar}
\email{kkma@physik.uni-bielefeld.de}

\author[0000-0003-4528-2745]{M.A. Krishnakumar}
\affiliation{Fakult{\"a}t f{\"u}r Physik, Universit{\"a}t Bielefeld, Postfach 100131, D-33501 Bielefeld, Germany}
\affiliation{Radio Astronomy Centre, NCRA-TIFR, Udagamandalam, India}
\affiliation{National Centre for Radio Astrophysics, Tata Institute of Fundamental 
Research, Pune, India}
\affiliation{Bharatiar University, Coimbatore, India}

\author[0000-0002-0862-6062]{Yogesh Maan}
\affiliation{ASTRON, Netherlands Institute for Radio Astronomy, Oude Hoogeveensedijk 4, 7991 PD, Dwingeloo, The Netherlands}

\author[0000-0002-0863-7781]{B.C. Joshi}
\affiliation{National Centre for Radio Astrophysics, Tata Institute of Fundamental 
Research, Pune, India}

\author[0000-0003-4274-211X]{P.K. Manoharan}
\affiliation{Radio Astronomy Centre, NCRA-TIFR, Udagamandalam, India}
\affiliation{National Centre for Radio Astrophysics, Tata Institute of Fundamental 
Research, Pune, India}

\begin{abstract}
We present multi-frequency scatter broadening evolution of 29 pulsars observed 
with the LOw Frequency ARray (LOFAR) and Long Wavelength Array (LWA). We conducted
new observations using LOFAR Low Band Antennae (LBA) as well as utilized the
archival data from LOFAR and LWA. This study has increased the total of all
multi-frequency or wide-band scattering measurements up to a dispersion measure
(DM) of 150~pc\,cm$^{-3}$ by 60\%. The scatter broadening timescale ($\tau_{sc}$) 
measurements at different frequencies are often combined by scaling them to a
common reference frequency of 1\,GHz. Using our data, we show that the
$\tau_{sc}$--DM variations are best fitted for reference frequencies close
to 200--300\,MHz, and scaling to higher or lower frequencies results in
significantly more scatter in data. We suggest that this effect might indicate
a frequency dependence of the scatter broadening scaling index ($\alpha$).
However, a selection bias due to our chosen observing frequencies can not
be ruled out with the current data set. Our data did not favour any particular model 
of the DM -- $\tau_{sc}$ relations, and we do not see a statistically significant
break at the low DM 
range in this relation. The turbulence spectral index ($\beta$) is found to be 
steeper than that is expected from a Kolmogorov spectrum. This indicates that the 
local ISM turbulence may have a low wave-number cutoff or presence of 
large scale inhomogeneities in the line of sight to some of the reported pulsars.
\end{abstract}


\section{Introduction} \label{intro}

Pulsed radio signals from pulsars get scattered due to multi-path propagation 
while traveling through the irregularities in the ionized inter-stellar medium (IISM), 
resulting in broadening of the pulse profile. This scatter broadening is a highly 
frequency dependent phenomenon, with the characteristic pulse broadening time, 
$\tau_{sc}$, scaling with the observing frequency, $\nu$, in the form of $\tau_{sc} 
\propto \nu^{-\alpha}$ \citep{rick77}, where $\alpha$ is the frequency scaling index. 
Consequently, the nearby pulsars (low DM\footnote{ DM: Dispersion Measure is the 
integrated column density of free electrons between the pulsar and the observer.}) 
show prominent scatter-broadening primarily 
at low radio frequencies ($<$ 1.0 GHz), while the distant pulsars (high DM) show 
measurable scattering effects up to much higher radio frequencies (a few GHz)  
and disappear completely as a pulsed source at low frequencies. The evolution of 
scatter-broadening of pulsar signal over frequencies provides insights into the 
turbulence characteristics of the IISM and allows a study of the clumps and hollows 
in the IISM in detail.

The multi-path propagation of the signal creates a diffraction pattern at the observer's 
plane that decorrelates over a characteristic bandwidth $\delta\nu_d$, such that 
2$\pi\tau_{sc} \delta\nu_d = C_1$. The constant $C_1$ is expected to be of the order of 
unity for a Kolmogorov type turbulence \citep{cwb85}.  Following \citet{rick77} and 
\citet{ars95}, the power spectrum of electron density irregularities for an isotropic 
homogeneous medium can be shown as

\begin{equation}
P_{n_e}(q) = C_{n_e}^{^2} (q^2 + l_o^{-2})^{-\beta/2} exp (-q^2 l_i^2 / 2),
\end{equation}
\noindent where $P_{n_e}(q)$ is the power spectrum density of irregularities, $C_{n_e}^{^2}$
is the scattering strength of the medium, $q$ is the three dimensional wavenumber, $\beta$
is the spectral index, $l_o$ and $l_i$ are the outer and inner scales of turbulence. This 
can be simplified further, if $l_i < 1/q < l_o $ to $P_{n_e}(q) = C^{2}_{n_e} q^{-\beta}$.
By combining different measurements of ISM effects, like dispersion, scattering, Faraday 
rotation, etc. on pulsars with distances $\leq$1~kpc, \citet{ars95} found that the electron
density turbulence is close to Kolmogorov model with an upper limit on the inner scale of 
turbulence as 10$^8$m. 
As mentioned above, $\tau_{sc}$ increases with decreasing observing frequency, and its 
frequency scaling index is related to $\beta$ via $\alpha = 2\beta/(\beta-2)$ (for $2 < 
\beta < 4$). For a Gaussian distribution of irregularities, it was found that $\tau_{sc} 
\propto \nu^{-4}$ DM$^{2}$ \citep{rick77}.
For the Kolmogorov turbulence model, one expects $\beta = 11/3$, and consequently 
the scaling relation changes to $\tau_{sc} \propto \nu^{-4.4}$ DM$^{2.2}$ \citep{rnb86}. 
\citet{gn85} worked out the relations for the cases where $\beta$ was found to be more 
than the critical value of 4. They speculated, with the then available data, that $\beta$ 
is 4.3 in general, instead of the 11/3 expected for a medium with Kolmogorov turbulence.



Several studies have been conducted in the past to understand the turbulence in the IISM 
by using different techniques. \citet{cwb85} studied the scintillation characteristics of 
a large sample of pulsars, and found that the IISM turbulence follows Kolmogorov turbulence 
very well in nearby regions ($\leq$ 2kpc). \citet{lkmll01,lmgka04} studied moderate to high 
DM pulsars with multi-frequency scatter-broadening measurements and proposed that the IISM 
tends to follow Kolmogorov turbulence up to about a DM of 350~pc~cm$^{-3}$ and deviates from 
it beyond this. \citet{lw3} analyzed a large sample of scatter-broadening measurements (60 
pulsars) to understand the characteristics of the IISM. They suggested that the IISM as a 
whole follows Kolmogorov turbulence characteristics. \citet{kbm17} undertook an  
extensive study of multi-frequency scatter-broadening of 39 pulsars using the Ooty Radio 
Telescope (ORT) and the Giant Metrewave Radio Telescope (GMRT), which increased the total 
number of available $\alpha$ measurements by about 50\%. They found that several nearby 
pulsars show a flatter $\alpha$ than expected. A study by \citet{gkk+17} on 
scatter-broadening evolution of 13 low DM pulsars using the LOw Frequency ARray (LOFAR) also 
showed that most of the nearby pulsars have a shallower $\alpha$ ($< 4$). They also noted that 
there may be a frequency dependent evolution of $\alpha$, after comparing their measurements 
with those at higher frequencies. One of the possibilities is that the pulsars in the 
supernova remnants or those with lines-of-sight passing through H{\small II} regions show a 
flatter $\alpha$ \citep{gn85}, however, this has not been explored in detail so far.

In all the earlier studies, very few measurements of scatter-broadening of low DM pulsars 
were included. This made it difficult to understand the turbulence characteristics in the 
local IISM. It has not been clear, with the sparse amount of data available so far, whether 
the turbulence characteristics are Kolmogorov or not  at low DMs. As explained above, it is 
expected that $\tau_{sc}$ $\propto$ DM$^{2.2}$ at low DMs and becomes steeper at high DMs. 
With the introduction of new low frequency telescopes with wide-band receivers and 
improved back-ends, we are now able to observe the in-band evolution of scatter-broadening 
of many low DM pulsars. In this study, we have made 
new observations as well as analyzed archival data of a large sample of low DM pulsars  
over multiple frequencies or by using data from large fractional bandwidth receivers 
to better understand the $\tau_{sc}-$DM dependence. With the advantage of having 
wide-band observations, we also measure and study in-band evolution of the frequency scaling index of scattering 
($\alpha$) for these pulsars.

In this study, we aim to (1) estimate $\alpha$ for a large set of pulsars below a DM of 
$\sim$150~pc~cm$^{-3}$, (2) confirm or rule-out previously reported dip in $\alpha-$DM 
relation around a DM of $\sim$100~pc~cm$^{-3}$ with better statistics, (3) model the DM$-\tau_{sc}$ 
relationship, and (4) examine for H{\small II} region or Nebula associations in the 
directions of the pulsars. The observations and analyses are described in Section \ref{obs} 
followed by our main results (Section \ref{res}). Section \ref{disc} presents a discussion 
of our results, and we summarize and conclude in Section \ref{conc}.

\section{Observations and Data Analysis}\label{obs}

\subsection{Observations and data reduction}

The data used in this paper were obtained from new observations with LOFAR as well as from 
the archived ones from both LOFAR and Long Wavelength Array (LWA). The LOFAR telescope has 
a seamless frequency coverage from $\sim$30 MHz to $\sim$240~MHz \citep{van13}. We have 
used the Low Band Array (LBA; 30$-$80\,MHz) to observe 31 pulsars. The observations were 
performed between December 2016 and May 2017 in several different sessions. The voltage beam 
data were coherently dedispersed by the observatory using the PulP pipeline on CEP3 cluster 
\citep{pulp} available at the facility. The dedispersed data were further processed using 
the PSRCHIVE package\footnote{PSRCHIVE website: http://psrchive.sourceforge.net/} \citep{psrchive} to excise radio frequency interference, correct for 
any possible slight inaccuracies in pulsar period and DM from the ephemerides and then 
reduced to sub-banded profiles for further processing. Although we observed 31 pulsars below 
a DM of 100~pc~cm$^{-3}$, processing of the data for pulsars with DMs above 40~pc~cm$^{-3}$ 
was not possible due to some technical issues with the then available version of the PulP. 
Additionally, there were other factors during our observations which reduced the overall 
achievable sensitivity\footnote{From our discussions with the observatory scientists, we gathered that LOFAR does not yet obtain full coherency in the LBA mode, lowering the sensitivity significantly. Although the issue is still under investigation, it is suspected that a better calibration of the instantaneous delays between the reference clock and all the stations is needed.}. We could detect 15 pulsars from our full list. Moreover, the signal-to-noise ratio (S/N) was not adequate enough for many of these to measure scatter broadening of the average profile or of those in separate parts of the band (for fitting purpose, we have used only profiles having a peak S/N ratio of at least 6 in this paper). We were able to measure $\tau_{sc}$ for 8 pulsars and estimate $\alpha$ for 6 of these using the sub-band data. There were not enough sub-band profiles with S/N ratio larger than 6 for the remaining two pulsars.

We used the LOFAR archival data reported in \citet{bilous16,pilia16} utilizing the High Band 
Array (HBA) in the frequency range of 110$-$190\,MHz. We re-processed the data using PSRCHIVE 
package to correct 
for any small inaccuracies in period and DM, and reduced these to sub-banded profiles. Depending on 
the sensitivity of detection, we used different number of sub-bands for each pulsar 
(typically 10 or 20) across the band. We measured $\tau_{sc}$ evolution across the band for 
a total of 14 pulsars combining data from other high frequency observations from the EPN 
data archive\footnote{European Pulsar Network archive for pulsar profiles at multiple 
frequencies. Website: http://www.epta.eu.org/epndb/} and from the ORT and GMRT (see KJM17
for the profiles and $\tau_{sc}$ estimates).

The LWA is a multipurpose radio telescope operating in the frequency range of 10 -- 88~MHz 
\citep[See][for more information about LWA]{lwa}. With the first station of LWA, namely 
LWA1, located near the Very Large Array (VLA) in New Mexico, \citet{srb15} reported the 
detection of 44 pulsars. These data are publicly available through the LWA 
data archive\footnote{http://lda10g.alliance.unm.edu/}. We chose a set of 10 pulsars from 
this archive, which show clear evolution of scatter-broadening and have peak S/N ratio 
larger than 6 even when the full band is divided in to 8 or 16 sub-bands for measuring 
$\tau_{sc}$. The data were reduced using PSRCHIVE package and analyzed further to measure $\tau_{sc}$.

Some of the pulsar profiles are scatter-broadened over the entire band of LOFAR and LWA. 
Finding a good template for the fitting procedure, detailed in the next session, was 
difficult from the archival data-sets for such pulsars. We observed a sub-set of pulsars 
from this list, which were within the declination coverage limits ($-60 > \delta < +60$) 
of the ORT. ORT is operating at a central frequency of 326.5~MHz with a bandwidth 
of 16~MHz. It is an ideal instrument to provide template profiles since frequency of 
observations of such templates 
will be close to the LWA and LOFAR frequencies and yet show no scatter-broadening.
Hence, the effects due to profile evolution will be limited, in contrast with measurements
using a higher frequency template. The observations were made using the PONDER back-end 
\citep{ponder} with the ORT during MJDs 57963 -- 57964. We detected 15 pulsars with good 
S/N ratio $(> 20)$ and used their profiles as templates for fitting. Out of the 15 detected pulsars, 
six were used as templates for HBA data, three each for LWA and LBA data and three for the 
data taken from LBA, HBA as well as LWA. For pulsars that were not detected (or were 
beyond the declination limits) at ORT, we used the nearest high frequency profiles from 
the EPN database. Most of these profiles were obtained from the Lovell telescope 
\citep{gl98} except one, which was from Arecibo \citep{snt97}, as mentioned in Table~\ref{tab1}.

\subsection{Analysis} \label{anal}

The average profiles, and in many cases the sub-band profiles, obtained from the above
analysis were further used for obtaining $\tau_{sc}$ by the method detailed in KJM17. 
Briefly, the scattered pulse profile can be represented as a convolution of the intrinsic 
pulse, $P_{i}(t)$, with the impulse response characterizing the scatter-broadening in the 
IISM, $s(t)$, the dispersion smear across the narrow spectral channel, D(t), and the 
instrumental impulse response, $i(t)$. The instrumental impulse response is small enough to 
neglect, since the rise times of the receivers and back-ends are very small. Hence, we have 
the scattered pulse as

\begin{equation}
P(t) = P_{i}(t) \ast s(t) \ast D(t)
\label{eq1}
\end{equation}

We followed the same method as detailed in KJM17 to extract 
$\tau_{sc}$, where we used a high frequency, unscattered profile for convolving with the 
impulse response of IISM, which can be expressed as

\begin{equation}
s(t) = \exp (-t/\tau_{sc}) U(t)
\label{eq2}
\end{equation}

where U(t) is a unit step function.

The data from LOFAR were coherently dedispersed, so the effect of $D(t)$ in Equation 
\ref{eq1} can be neglected. LWA data on the other hand were incoherently dedispersed, with 
a specific channel resolution. This introduces an increase in the observed pulse profile 
width. We convolved the template profiles with the estimated dispersion smearing before 
fitting for scatter-broadening in such cases, to minimize any contamination of the resultant 
measurements.

After measuring $\tau_{sc}$ for each of the sub-banded profiles, we used a Monte-Carlo 
method described in KJM17 for estimating the frequency scaling index, $\alpha$.  For 
$\tau_{sc}$ estimates at each of the sub-bands, we generated 10,000 normally distributed 
random numbers with the root-mean-square deviation same as the error bars on $\tau_{sc}$.
To estimate $\alpha$, a power-law fit was made to these randomly generated $\tau_{sc}$ values at 
different frequencies. We report the median of these $\alpha$ values, with error bars from 5 
and 95 percentiles, in Table~\ref{tab1}.  Some of the pulse profiles 
at the lower frequency end look very noisy, mostly because of the large number of bins which
is kept same for all the sub-bands.
The uncertainties are naturally large for the $\tau_{sc}$ measurements of such sub-band profiles.
The $\alpha$ measurements take in to account the uncertainties on $\tau$, and hence, are
not affected by such low S/N profiles. In any case, we confirmed that the final value of
$\alpha$ did not show any deviations beyond the limits yielded from
the Monte-Carlo method, when we tried removing those noisy lower-frequency profiles
from the analysis.

\begin{figure}
\centering
\begin{minipage}[b]{0.5\textwidth}
\includegraphics[scale=0.5, angle=0.0]{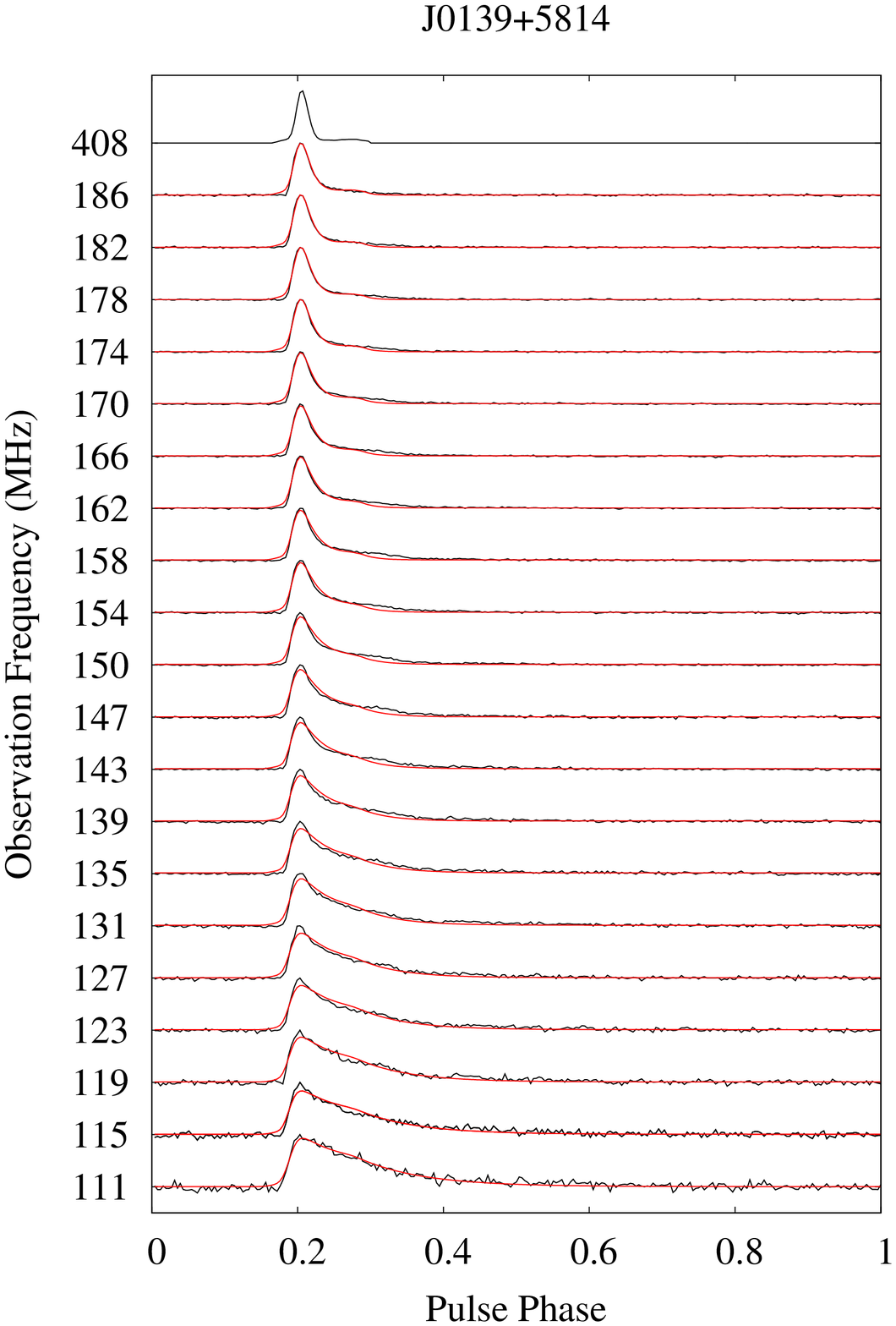}
\end{minipage}
\hfill
\begin{minipage}[b]{0.4\textwidth}
\hspace{-0.2in}\includegraphics[scale=1.1, angle=0]{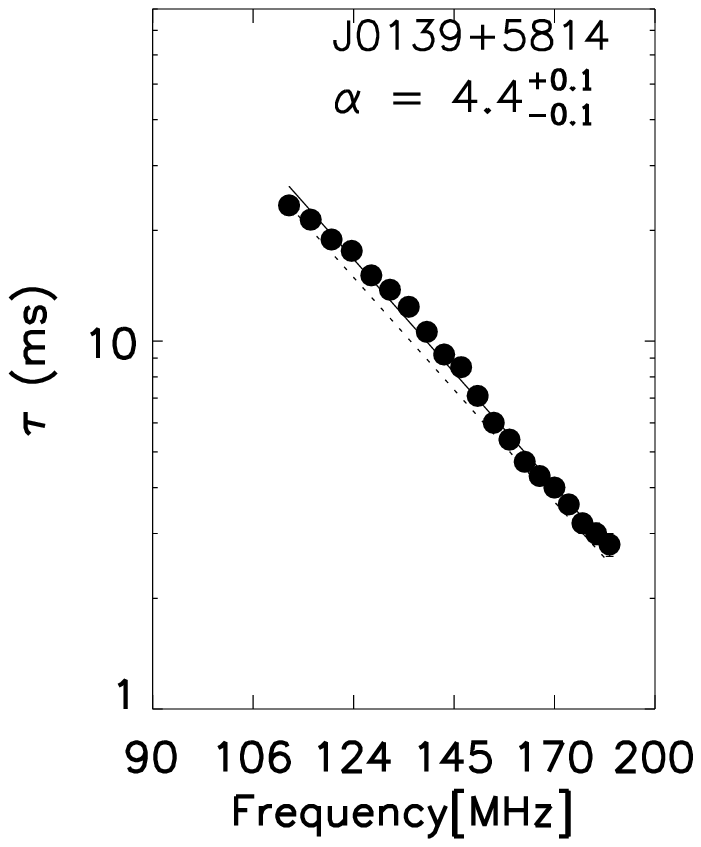}
\end{minipage}
\caption{Frequency evolution of scatter broadening for PSR J0139+5814. \textbf{Left panel} shows
the pulse profiles at different frequencies as labeled in the abscissa. The 
profile at the top is the template profile used for fitting purpose to estimate $\tau_{sc}$. 
The black color curves are observed profiles at different frequencies and the red curves
are the best fit models. \textbf{Right panel} shows $\tau_{sc}$ as a function of observing
frequency. The black filled circles are the measured 
$\tau_{sc}$ and the black continuous line is the fit to the complete data. The dashed line 
represents the trend corresponding to the typical Kolmogorov spectrum value of 4.4. Errorbars
on $\tau_{sc}$ are smaller than the symbol sizes.}
\label{samplefig}
\end{figure}

\section{Results} \label{res}

We report scatter-broadening measurements of 31 pulsars, and $\alpha$ estimates for 29 of 
these. The results are summarized in Table~\ref{tab1}, with details like, pulsar name in J2000
coordinates, period, DM, distance, number of bins in the profiles, range of frequencies from
which $\tau_{sc}$ is estimated, reference template frequency, and
value of $\alpha$ and $\tau_{sc}$ scaled to 200 MHz. The last column lists the telescopes
used in this study, in the ascending order of frequencies, with the highest frequency
providing the template obtained from the telescope mentioned last\footnote{A complete set of
$\tau_{sc}$ measurements and plots of sub-band profiles used, along with fits (similar to
Figure~\ref{samplefig}) for each pulsar can be found in the supplementary material and is also
available in  electronic form at http://rac.ncra.tifr.res.in/data/pulsar/supplementary-material-KYBM.pdf}. A sample plot 
of the scatter broadening evolution of PSR J0139+5814 is shown in Figure~\ref{samplefig}. 
The left hand side panel shows the sub-band integrated pulse profiles at different frequencies 
obtained using the HBA of LOFAR, where one can see a clear evolution of scatter broadening 
as a function of frequency. We have used a high frequency profile as a template 
for estimating the scatter broadening, which, in this case, is the profile at 408 
MHz from the Lovell telescope, as shown at the top of the panel. The right hand side panel 
shows fit for estimating $\alpha$. No measurement of $\alpha$ was possible for pulsars 
where $\tau_{sc}$ could be estimated for only two or less frequencies. For these pulsars, 
estimates of $\tau_{sc}$, scaled to 200 MHz by using $\alpha = 4$ are also listed in 
Table~\ref{tab1}. Out of the 31 pulsars, 14 are 
from the HBA archival data, 7 are from the LWA archival data, 7 are from our observations 
with the LBA, an additional one is common between both the LWA archive and the LBA 
observations, and another two are common between both the LWA and the HBA archives. This 
study has increased the total set of $\alpha$ measurements below a DM of 
$\sim$100~pc~cm$^{-3}$ by two folds. The $\alpha$ value estimated in this study ranges between 
2.4$-$5.7. We note that several pulsars in this DM range show a flatter $\alpha$ than what 
was reported previously by \citet{cwb85} and \citet{jnk98}, using decorrelation bandwidth
measurements. This can also be seen in the recent results from \citet{gkk+17} and KJM17.

Five of our pulsars are common to \citet{gkk+17}, viz., PSRs J0040+5716, J0117+5914, 
J0543+2329, J1851+1259 and J1913$-$0440. \citet{gkk+17} measured scatter broadening 
by simultaneously fitting for both $\tau_{sc}$ and the intrinsic profile width, $P_i(t)$.
Since the intrinsic pulse width is unknown and is also found to increase with
decreasing observing frequency in general\citep{thor91,pilia16}, a fit to both these 
parameters make it difficult to decouple scattering effects from profile evolution. In 
addition, the pulse broadening functions used in \citet{gkk+17} are different, hence our 
results cannot be directly compared with those reported by them. While profile evolution
can also affect our analysis, making the lowest frequency $\tau_{sc}$ measurement an over
estimate, we used unscattered profile at a nearby frequency (mostly with ORT at 327 MHz)
to mitigate this effect.  For a fair comparison, we 
reanalyzed the data for these five pulsars using \citet{gkk+17} methods. We find that the 
measurements of $\alpha$ are consistent with their results within error bars for  PSRs J0040+5716, J0117+5914, 
J0543+2329 and  J1851+1259 with both their anisotropic and isotropic models. In case of 
PSR J1913$-$0440, our $\alpha$ is  steeper than that from both of their models by about
three times the standard 
deviation. However, this is perhaps due to the fact that we did not have access to data 
below 148 MHz for this pulsar. Thus, our estimates of $\alpha$ from independent measurements 
of $\tau_{sc}$ are consistent with the previously reported values.

There are six pulsars in our data-set where the profiles have multiple components. Two of
them (PSRs J0406+6138 and J0629+2415) show the steepest $\alpha$ in our list, while the rest 
four (PSRs J0139+5814, J0525+1115, J2149+6329 and 2229+6205) exhibit flatter $\alpha$. It is 
not clear whether profile evolution has affected these measurements, but we do not see any 
significant profile evolution between the template profile and the highest frequency profile 
in the HBA band, where $\tau_{sc}$ is measured. Although there is large amount of scatter 
broadening for three of them, the two pulsars with steep $\alpha$ have $\tau_{sc}$ less than  
10\% of the period at the lowest frequency in HBA. Hence, to clearly understand whether our 
measurements are affected by profile evolution, we require lower frequency observations 
(LWA/LBA frequencies), where the scatter-broadening will dominate the intrinsic pulse width 
evolution. In case of the four flatter $\alpha$ pulsars, except for PSR J0139+5814, the other 
three are heavily scatter-broadened. Hence, it is unlikely that the intrinsic profile width 
evolution had any significant effect on the measured $\alpha$.

Two pulsars (PSRs J1645--0317 and J1752--2806) in our sample have previous $\alpha$ 
measurements available in the literature \citep{cwb85} and two others (PSRs 
J0358+5413 and J0826+2637) have multi-frequency  decorrelation bandwidth ($\delta\nu_d$)
estimates. \citet{cwb85} 
report $\alpha$ of $4.5\pm0.4$ and $4.7\pm0.3$ for PSR J1645--0317 and PSR J1752--2806,
respectively, using $\delta\nu_d$ and $\tau_{sc}$ estimates made at discrete frequencies 
between 80 and 1400 MHz. They had only one low frequency $\tau_{sc}$ estimate (at 80 MHz) 
while all other measurements were $\delta\nu_d$ estimates from higher frequency observations.
With the data presented here, our estimates of $\alpha$ are 4.1$\pm$0.6 and 3.7$\pm$0.2 
respectively for these pulsars. It can be seen that $\alpha$ is different for PSR J1752--2806 
while it is 
consistent within limits for PSR J1645--0317. The change in $\alpha$ for PSR J1752--2806 might 
indicate the possibility of its frequency dependence, if it is not caused by an erroneous value of $C_1$
or due to changes in the IISM turbulence characteristics over several years.
For the other two pulsars, we found 
$\delta\nu_d$ measurements at higher frequencies ($>$ 400~MHz) 
from \citet{cwb85}. The $\alpha$ values obtained after combining with our data set are different 
than what is reported here ($\alpha=3.5\pm0.1$ for PSR J0358+5413 and $\alpha=4.8\pm0.1$ for PSR 
J0826+2637). Without a good understanding of the actual value of $C_1$, combining the 
$\delta\nu_d$ data with $\tau_{sc}$ measurements might give incorrect results. Since we do not 
have the measured 
value of $C_1$, we used the theoretical value of unity, for demonstrating various possibilities 
that may affect the determination of $\alpha$. The changes in the combined data of the latter 
pulsars may be due to either a frequency dependence of $\alpha$ or as in the previous case, due 
to the slow long-term variations in the IISM turbulence. We are unable to make any claim of a
frequency dependence of $\alpha$ on the basis of this data combination, due to the unavailability 
of simultaneous multi-frequency data across a wide range of frequencies and the lack of knowledge 
of the actual value of $C_1$, while converting from $\delta\nu_d$ to $\tau_{sc}$. 

With this study, the total number of $\alpha$ measurements have increased to 111. Further, 
below a DM of 200 pc~cm$^{-3}$, this number has increased to 83, around 50\% increase from 
the sample that was available with KJM17. A dip at around a DM of 100~pc~cm$^{-3}$ was 
reported by KJM17 as well as by \citet{gkk+17}. As discussed in the following section, we do 
not see this dip in our now statistically large sample.

\section{Discussions} \label{disc}

\subsection{DM dependence of scattering}

\citet{rick77} suggested that there is probably a power-law relation between DM and 
$\delta\nu_{d}$, although the number of available 
measurements were very less at that time. If the scattering material is distributed uniformly along the 
distance to the pulsar, then $\tau_{sc} \propto$ DM$^{2}$ and for a power-law turbulence with 
Kolmogorov spectrum characteristics, $\tau_{sc} \propto$ DM$^{2.2}$. Later, \citet{slee80} extended 
the sample with scintillation and scatter-broadening measurements of 31 pulsars and reported 
a power-law relation between DM and $\tau_{sc}$ with a power-law index of 3.5$\pm$0.2. 
\citet{asb86} found a similar relation ($\tau_{sc} \propto $DM$^{3.3\pm0.2}$) using a sample 
of 85 pulsars.  These results indicated that the $\tau_{sc}$ -- DM relationship is much 
steeper than what was expected from a Kolmogorov spectrum of turbulence and requires a 
higher $\beta$. \citet{gn85} realized this problem and suggested that this can be 
accommodated by theory within the range of $4<\beta<6$. They also argued that this may be 
due to scattering from large scale inhomogeneities in the line of sight to the pulsar.

\citet{bwhv92} fitted a broken power-law (sum of two power-laws) to the $\tau_{sc} - $DM 
relation, since they saw a flattening from a simple power-law spectrum at low DMs (below
a DM of $\sim$30 pc~cm$^{-3}$). Their fit 
showed that the measured $\tau_{sc}$ follows Kolmogorov turbulence at low DMs and deviates 
at higher DMs. This was adapted in many later works \citep{rm97}. Similarly, \citet{bhat04} 
used a simple parabolic curve  
for fitting 371 $\tau_{sc}$ measurements at different frequencies and estimated $\alpha$ to be 
3.86$\pm$0.16, i.e., much less than 4.4 expected for a medium with Kolmogorov turbulence. In these 
studies, $\tau_{sc}$ was scaled to a reference frequency (typically 1 GHz) assuming a 
constant $\alpha$. In contrast, \citet{lw3} scaled the $\tau_{sc}$ by using the $\alpha$ 
obtained from the measurements till then (60 pulsars) to 1 GHz, before studying its 
dependence with DM. With the new $\alpha$ estimates from KJM17, \citet{gkk+17}, and those 
from this study, which has increased the sample by about 2 folds (111 pulsars) from what was 
available with \citet{lw3}, we try to re-examine these relations.

Available measurements of $\alpha$ typically have 5$-$10$\%$ uncertainties. Thus, scaling 
$\tau_{sc}$ from the frequency of its estimate to a widely different frequency is likely to
not only yield an over-estimate of uncertainty at the reference frequency, but also an 
increased scatter. Thus, high frequency measurements of $\tau_{sc}$  will yield large 
scatter at low reference frequency and vice-verse.  While \cite{kmnjm15} estimated 
$\tau_{sc}$ for a large number of pulsars at a single frequency of 326.5 MHz, this is not 
possible for the pulsar population as a whole as  $\tau_{sc} - $DM relation implies 
significant scattering for low DM pulsars at frequencies only below 300 MHz and reliable 
measurements for high DM pulsars possible only for frequencies above 1 GHz. Thus, an optimum 
reference frequency needs to be chosen for comparison of variations in $\tau_{sc}$ as a 
function of DM.

\begin{figure}[t]
\centering
\includegraphics[scale=0.5, angle=-90]{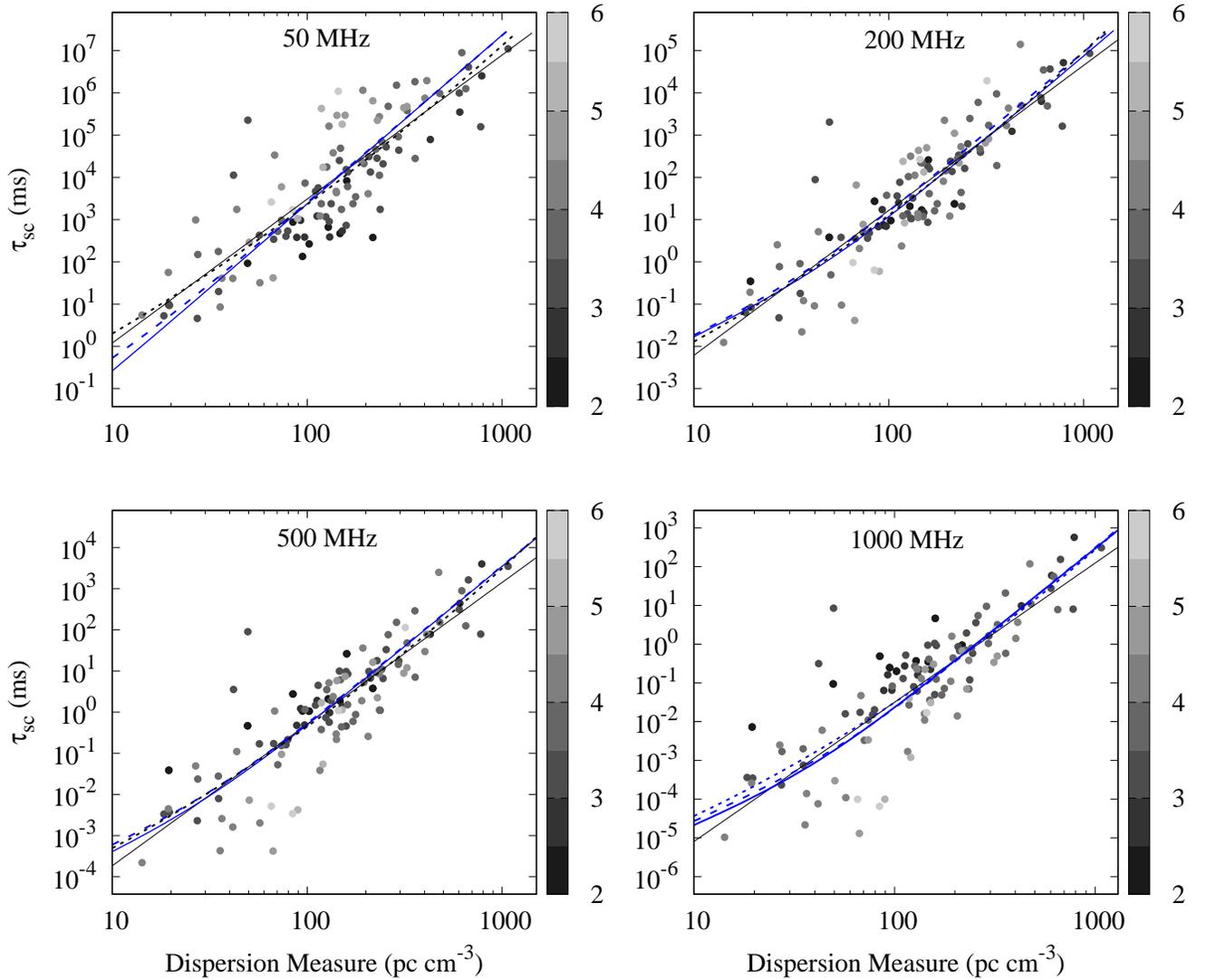}
\caption{The panels of the figure show the scaled $\tau_{sc}$ against DM plotted at 50, 200
500~MHz and 1~GHz. The blue continuous and dashed curves represent the fit of broken power 
laws, as given in Equation~\ref{eq11}. Black dashed curve is the fit of a parabolic function 
following the relation in Equation~\ref{eq12} and black continuous line is the fit by a simple 
power-law. The $\tau_{sc}$ measurements are plotted with a color scheme of light gray to 
dark gray as a function of their $\alpha$ value in steps of 0.5 between $2 < \alpha < 6$. The 
data used in this plot are taken from \citet{lkmll01,lmgka04}, \citet{lw1,lw2,lw3}, \citet{gkk+17}, 
KJM17 and this study.}
\label{tdmplot}
\end{figure}

\begin{figure}[t]
\centering
\includegraphics[scale=0.5,angle=-90]{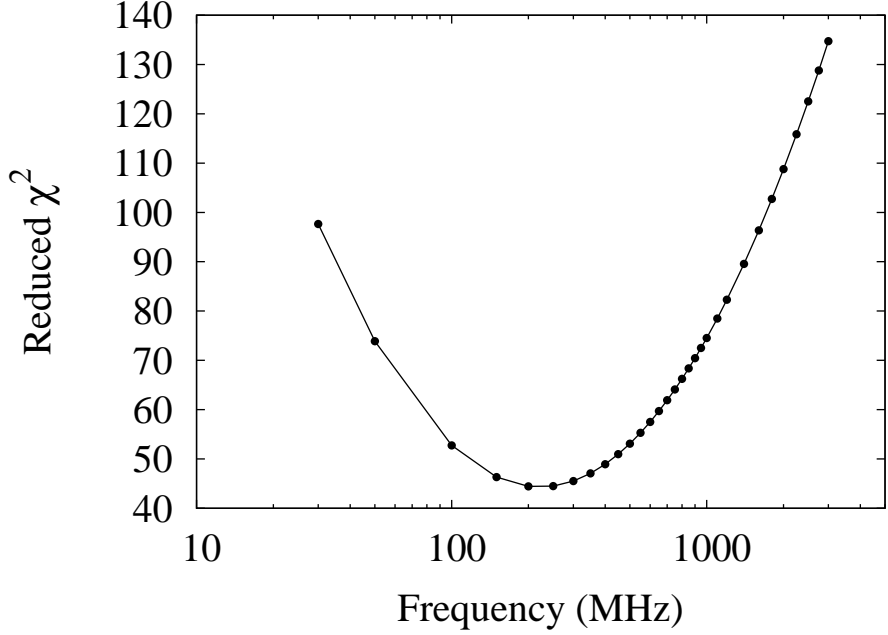}
\caption{The plot shows the reduced $\chi^{2}$ of a simple power-law fit to the whole data 
set of $\tau_{sc}$ against DM as a function of reference frequency. The black points denote 
the reduced $\chi^{2}$ at each frequency where a fit was performed.}
\label{chifrq}
\end{figure}

Before scaling $\tau_{sc}$ to a standard frequency of 1 GHz, it is worthwhile to check 
if 1 GHz is the optimum frequency for scaling all these measurements, since almost all of 
the $\tau_{sc}$ measurements were obtained below 1 GHz. We first chose a set of trial 
reference frequencies between 30 and 1000~MHz, each separated by 50~MHz. A plot of scaled 
$\tau_{sc}$ at 50~MHz, 200 MHz 
500~MHz and 1~GHz are given in Figure~\ref{tdmplot} for comparison. Then, for each of 
the trial reference frequencies, we  scaled $\tau_{sc}$ using the best fit parameters
obtained while fitting for $\alpha$. We fitted a simple power-law as was done by 
\citet{slee80,asb86} and computed reduced $\chi^{2}$ of the fit. Although the $\chi^{2}$ of 
the fits are very high, it does vary significantly with chosen reference frequency for 
scaling, as shown in Figure~\ref{chifrq}. It is clear that the minimum of 
the $\chi^{2}$ for our data set is around 200\,MHz, indicating that the scatter in the plot 
is minimized at this frequency. Thus, we chose
200 MHz as a reference frequency for 
comparison of different models. The model relations we chose from the literature
are (1) a simple power law model \citep{slee80,asb86}, (2) a broken power law model as in 
Equation~\ref{eq11}, (3) a broken power law model by fixing the value of $\zeta = 2.2$ in
Equation~\ref{eq11} \citep{bwhv92,rm97}, and (4) a second order polynomial fit following 
\citet{bhat04} as given
in Equation~\ref{eq12}. Since we measured $\alpha$ for the pulsars used in this study, we
have scaled $\tau_{sc}$ using it and removed the last term in Equation~\ref{eq12} while fitting
\citep{lw2,gkk+17}.

\begin{equation}
\tau_{sc} = a DM^{\zeta} ({1 + b DM^{\gamma}})
\label{eq11}
\end{equation}

\begin{equation}
log \tau_{sc} = a + b (log DM) + c (log DM)^{2} - \alpha log \nu
\label{eq12}
\end{equation}

The power law fit to the 200 MHz data yielded $\tau_{sc} = DM^{3.4} / 10^{5.6}$, with a 
reduced $\chi^2$ of 44. The broken power law model, following Equation~\ref{eq11} gave
$\tau_{sc} = 1.09 \times 10^{-4} DM^{2.2} ({1 + 2.7 \times 10^{-3} DM^{1.65}})$ with a 
reduced $\chi^2$ of 31. The broken power law model with the value of $\zeta$ fixed at 2.2
gave $\tau_{sc} = 9.95 \times 10^{-5} DM^{2.2} ({1 + 1.5 \times 10^{-3} DM^{1.7}})$ with
a reduced $\chi^2$ of 29. The second order polynomial, following Equation~\ref{eq12} (after
removing the last term of $\alpha$) yielded $log \tau_{sc} = -4.01 + 1.68 (log DM) + 0.45 
(log DM)^{2}$ with a reduced $\chi^2$ of 37. Above fits do not really 
favour any 
particular relations. The difference between various fitted forms will be more apparent at 
very low or high DMs. The current data set is insufficient to examine this due to the 
unavailability of a large number of $\alpha$ of pulsars below a DM of 30~pc~cm$^{-3}$ and 
no $\alpha$ measurements below a DM of 10~pc~cm$^{-3}$ from scatter broadening measurements. 
Thus, populating this DM range with new $\alpha$ estimates from telescopes with wide-band 
receivers, such as LWA, LOFAR and future Square Kilometer Array, will be helpful in 
clarifying this issue.

Our fit to Equation~\ref{eq12} shows that there is a strong linear dependence in 
the DM -- $\tau_{sc}$ relation than what was reported by \citet{bhat04}. \citet{lw3} and 
\citet{gkk+17} also reported a strong dependence on the linear term than the quadratic one. 
We also find that the scatter in our plots vary as a function of DM (as also demonstrated 
above using Figure~\ref{tdmplot}). At very low frequencies, i.e., below $\sim$100~MHz, 
although the $\tau_{sc}$ of low DM pulsars seem cluster together, the high DM ones are 
scattered across a large range. Likewise, at high frequencies, the scatter is high for 
low DM pulsars but is less for high DM pulsars. This change of scatter may be due to the 
frequency dependence of $\alpha$ or due to the fact that the $\tau_{sc}$ measurements of 
these pulsars were done at around these frequencies. Although there can be other reasons 
for scatter in measurements of $\tau_{sc}$, such as a low wave number cut-off in the 
turbulence spectrum or a non-Kolmogorov type turbulence in the line of sight, our analysis 
does indicate that there is a possibility for a frequency dependence of $\alpha$. Further 
wide-band multi-frequency observations are required to confirm or disprove this effect.

\subsection{Turbulence characteristics of the IISM}

\begin{figure}[t]
\centering
\includegraphics[width=\linewidth,angle=270.,scale=0.6]{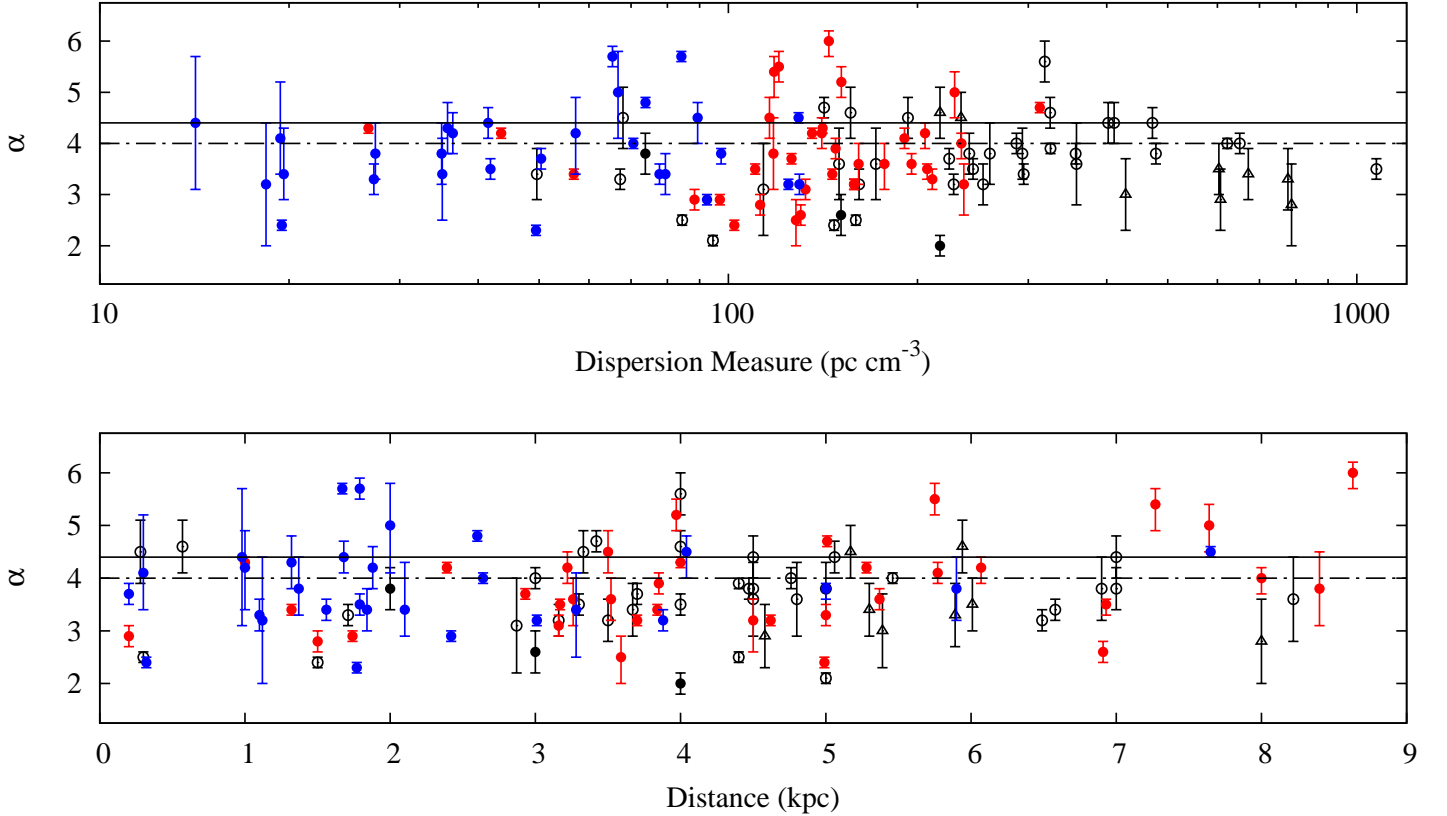}
\caption{The figure shows the distribution of $\alpha$ as a function of DM and distance. 
Black open circles denote measurements from \citet{lw1,lw2,lw3}; black filled circles are 
measurements from \citet{gkk+17}; black open triangles are from \citet{lkmll01,lmgka04}; red 
filled circles denote measurements from KJM17 and blue filled circles are from this study. 
The black solid line indicate $\alpha$ = 4.4 and black dash-dotted line shows $\alpha$ = 4.}

\label{dmdist}
\end{figure}

A plot of $\alpha$ against DM and distance to the pulsar are given in the top and bottom
panels of Figure~\ref{dmdist}. The plot is now mainly populated by measurements 
from this study and KJM17 up to a 
DM of $\sim$300~pc~cm$^{-3}$. The $\alpha$ measurements are scattered across a large range, 
but mostly below the conventional value of 4 or 4.4, at all DM and distance ranges. Unlike 
what was reported in KJM17, we do not see any clear dip in $\alpha$ at around a DM of 
100~pc~cm$^{-3}$. A closer inspection of the data from KJM17 and this study show that most 
of the pulsars with low $\alpha$ values around 100~pc~cm$^{-3}$ are concentrated behind the 
Sagittarius arm and some are behind H{\small II} regions or supernova remnants. A list of 
pulsars behind the H{\small II} regions and supernova remnants is given in Table~\ref{tab3}. 
The table lists, against each pulsar, the DM, distance estimated from the DM, 
distance estimated from parallax measurements, H{\small II} region along the line of sight 
to the pulsar and the distance to it and the $\alpha$ estimates. It should be noted that 
only some of the pulsars listed in the table show a flatter $\alpha$ than the average value.
Few low DM pulsars, without any associations, also show a low value of $\alpha$, as can be
realized from Table~\ref{tab1} , which may be due to a low wave-number cut-off in the 
turbulence spectrum at our low observing frequencies, affecting the measured $\alpha$ values 
as put forward by \citet{gn85}, or due to anisotropic scattering screens \citep{gkk+17}.
A detailed investigation of these effects is beyond the scope of the present work.

The turbulence spectral index, $\beta$, can be estimated from $\alpha$, for a power-law 
wave-number spectrum using the relations put forward by \citet{gn85}, \citet{cpl86} and 
\citet{rnb86} as given below

\begin{equation}
\alpha = 
  \begin{cases}
    \frac{2\beta}{\beta-2}    & \quad  2<\beta<4\\
    \frac{8}{6-\beta}         & \quad  4<\beta<6.
  \end{cases}
\label{eq6}
\end{equation}
\noindent The two regions of $\beta$ differ essentially due to the difference in the
scale sizes of the scattering regions involved; for $\beta<4$, the scattering comes from
the small scales($10^{9} - 10^{12}$ cm), while for $\beta>4$, the scales involved are much 
larger ($10^{12} - 10^{15}$ cm). In addition, for the $\beta>4$ region, the effect of
refractive scattering is large and is independent of observing frequency and distance to
the pulsar \citep{gn85}.

The distribution of the $\alpha$ values is not normal, and has a skewness of 0.3. Hence, 
estimating the average value of $\alpha$ might give wrong estimate. Hence, we report the 
median of this distribution with limits from the lower and higher quartiles as 3.7$^{+0.6}_{-0.7}$.
A similar distribution for both the relations in equation~\ref{eq6}, is also calculated, only with
larger skewness (mostly due to the effect of large/small values of $\beta$ near the boundaries) 
which makes median statistics a better estimate. The median value of this distribution of 
$\beta$ values came to be 4.2$^{+0.9}_{-0.6}$ and 3.8$\pm$0.3 respectively, for both the models.
\citet{gn85} assumed $\beta$ to be 4.3 for their discussions and were supported 
by observations from other fields also (see \citet{gn85} for more details). As is evident 
from our data-set, $\beta$ is much higher than $11/3$ and indicates that the turbulence spectra 
is steeper than the Kolmogorov model with a small inner scale. Following \citet{ars95} we 
estimated the power spectral density of scattering for all of the reported pulsars reported 
here and almost all the points followed the slope of the power spectrum with an inner scale ($l_i$) 
of 10$^5$~m and an outer scale ($l_o$) of 10$^{18}$~m, indicating that the effects of innerscale 
of turbulence is hard to decipher with the current data set and will require future observations.

\section{Summary and Conclusions}\label{conc}

We have reported multi-frequency scatter-broadening measurements of 31 pulsars, along
with frequency scaling index ($\alpha$) for 29 of these. This sample increases the 
number of the $\alpha$ measurements so far by about 30\% and almost doubles 
the sample below a DM of $\sim$100~pc~cm$^{-3}$. With this enriched data-set, we analyze the 
DM dependence of scattering. We find that the scatter in the $\tau_{sc}$ -- DM dependence 
changes as a function of reference frequency and hence scaling all the measurements to a 
standard frequency of 1~GHz, as was done in earlier studies, is not a good practice. This
analysis
also indicates that there is a possibility for $\alpha$ to be frequency-dependent, in 
addition to other effects which may cause this scatter. We tried to fit a 
power-law after scaling all the 111 measurements till date to different reference 
frequencies and find that around 200 -- 300~MHz is the best range for fitting the DM -- 
$\tau_{sc}$ relations, since the scatter in the data is minimal at those reference
frequencies. Furthermore, we 
find no significant difference between fits to different $\tau_{sc}$ -- DM relations in the 
literature, although a broken power-law fit seems to give the smallest reduced $\chi^{2}$. 
We find that $\alpha$ for many of the pulsars at low DMs in the current sample is 
consistently lower than the standard value of 4.4 for a Kolmogorov spectrum. This indicate 
that there may be a wave-number cut-off in these lines of sight or the presence of 
inhomogeneities like H{\small II} regions, stellar winds, supernova remnants, etc. 
The median value of $\alpha$ was found to be 3.7$^{+0.6}_{-0.7}$. We have also 
estimated the turbulence spectral index, $\beta$ for all the pulsars and found that the 
median value is 4.2$^{+0.9}_{-0.6}$ and 3.8$\pm$0.3 for the expressions of $\alpha$
given in Equation~\ref{eq6}. These values are 
higher than what is expected for a Kolmogorov spectrum ($\beta$=11/3) and is 
consistent with results from scattering in other fields \citep{gn85}. We did not find any 
significant evidence for a flattening of $\alpha$ around a DM of $\sim$100~pc~cm$^{-3}$ 
which indicates that this inference was perhaps biased by the sample selection. We also find that 
line of sight to some of the pulsars in the sample pass through H{\small II} regions or 
supernova remnants. Many of these pulsars show a flatter $\alpha$ than 4.4, indicating 
that the large electron number density, and hence turbulence of these regions, might affect 
scatter-broadening.

Further investigations are required to study the frequency dependence of $\alpha$, where low 
frequency telescopes with wide-band receivers such as LOFAR, LWA and SKA would play a 
vital role. Multi-frequency observations of a carefully chosen set pulsars will help further 
exploring any frequency dependence of $\alpha$ and we are in the process of conducting such 
observations in near future.


{\it Acknowledgment}: We thank the anonymous referee for detailed suggestions which helped 
in improving the manuscript substantially. We are grateful to Annya Bilous and Maura Pilia for 
sharing partially 
reduced archival LOFAR data with us. YM acknowledges use of the funding from the European 
Research Council under the European Union's Seventh Framework Programme (FP/2007-2013)/ERC 
Grant Agreement no. 617199. This paper is based (in part) on data obtained with the 
International LOFAR Telescope (ILT) under project code {LC7\_029}. LOFAR \citep{van13} is 
the Low Frequency Array designed and constructed by ASTRON. It has observing, data 
processing, and data storage facilities in several countries, that are owned by various 
parties (each with their own funding sources), and that are collectively operated by the ILT 
foundation under a joint scientific policy. We thank the scientists at LOFAR, particularly 
Sander ter Veen, for the support with our observations and processing on CEP3. We thank the 
LWA consortium for making their data publicly available. Support for operations and 
development of the LWA1 is provided by the National Science Foundation of the University 
Radio Observatory program. Basic research on pulsars at NRL is supported by the Chief 
of Naval Research (CNR). We also thank the technicians at the ORT for their support during 
the observations. We acknowledge support from Department of Science and Technology grant 
DST-SERB Extra-mural grant EMR/2015/000515.

\clearpage
\startlongtable
\begin{deluxetable}{clclcccccl}
\tablecolumns{10}
\tabletypesize{\scriptsize}
\tablecaption{Measurements of $\alpha$ for pulsars from our analysis. Each column gives 
pulsar name in J2000, Period in seconds, DM in pc~cm$^{-3}$, Distance derived from YMW16 
model \citep{ymw16}, Number of phase bins used in the profile, Frequency range of $\tau_{sc}$ 
measurements, Template profile frequency used for fitting, $\alpha$ estimated, $\tau_{sc}$ 
scaled to 200~MHz and the telescope(s) used. An asterisk with the distance estimate show 
that it is a DM independent distance estimate. For pulsars which do not have $\alpha$
estimates, the $\tau_{sc}$ is scaled assuming $\alpha = 4$ and are marked by a $\dagger$.
\label{tab1}}
\tablehead{
PSR J & Period &      DM      & Dist. & Bins & Freq. range & Templ. Freq. & $\alpha$ & $\tau_{sc}$ & Telescopes \\
      & (s)    & pc~cm$^{-3}$ &   (kpc)  &             &    (MHz)    &    (MHz)     &          &   (ms)   &           \\
}
\startdata
J0040+5716            & 1.118225 & 92.51  & 2.42  & 512  & 111.8 -- 186.1  & 408 & 2.9$^{+0.1}_{-0.1}$ & 17 $\pm$ 1          & HBA; LOVELL\\
J0117+5914            & 0.101439 & 49.42  & 1.77  & 256  & 111.8 -- 186.1  & 610 & 2.2$^{+0.1}_{-0.1}$ & 3.5 $\pm$ 0.2       & HBA; LOVELL\\
J0139+5814            & 0.272451 & 73.81  & 2.60* & 256  & 111.8 -- 143.1  & 408 & 3.1$^{+0.1}_{-0.1}$ & 1.9 $\pm$ 0.1       & HBA; LOVELL\\
J0358+5413            & 0.156384 & 57.14  & 1.00* & 256  & 47.35 -- 81.62  & 327 & 4.6$^{+0.7}_{-0.8}$ & 0.1 $\pm$ 0.1       & LWA; LOVELL\\
J0406+6138            & 0.594576 & 65.41  & 1.79  & 256  & 111.8 -- 158.7  & 408 & 5.7$^{+0.2}_{-0.2}$ & 1.0 $\pm$ 0.1       & HBA; LOVELL\\
J0415+6954            & 0.390715 & 27.45  & 1.37  & 512  & 115.7 -- 143.1  & 327 & 3.8$^{+0.6}_{-0.5}$ & 0.7 $\pm$ 0.2       & HBA; ORT\\
J0502+4654            & 0.638565 & 41.83  & 1.32  & 128  & 145.0 -- 408.0  & 610 & 3.5$^{+0.2}_{-0.2}$ & 100 $\pm$ 50        & HBA; GMRT; ORT; LOVELL\\
J0525+1115            & 0.354438 & 79.42  & 1.84  & 256  & 113.8 -- 152.8  & 327 & 3.4$^{+0.4}_{-0.4}$ & 4.4 $\pm$ 0.8       & HBA; ORT\\
J0543+2329            & 0.245975 & 77.70  & 1.56  & 256  & 113.8 -- 176.3  & 327 & 3.4$^{+0.2}_{-0.2}$ & 2.0 $\pm$ 0.3       & HBA; ORT\\
J0629+2415            & 0.476623 & 84.18  & 1.67  & 512  & 111.8 -- 154.9  & 327 & 5.7$^{+0.1}_{-0.1}$ & 0.54 $\pm$ 0.03     & HBA; ORT\\
J0826+2637            & 0.530661 & 19.48  & 0.32* & 512  & 32.65 -- 62.05  & 327 & 2.4$^{+0.1}_{-0.1}$ & 0.06 $\pm$ 0.01     & LWA; ORT\\
J0922+0638            & 0.430627 & 27.30  & 1.10* & 512  & 32.81 -- 47.46  & 327 & 3.3$^{+0.3}_{-0.3}$ & 0.05 $\pm$ 0.02     & LBA; ORT\\
J1509+5531            & 0.739682 & 19.62  & 2.10* & 512  & 37.61 -- 71.87  & 150 & 3.4$^{+0.9}_{-0.5}$ & 0.1 $\pm$ 0.1       & LBA; HBA\\
J1543$-$0620          & 0.709064 & 18.38  & 1.12  & 256  & 42.58 -- 57.22  & 150 & 3.2$^{+1.2}_{-1.2}$ & 0.1 $\pm$ 0.1       & LBA; HBA\\
J1543+0929            & 0.748448 & 34.98  & 5.90* & 256  & 52.25 -- 85.93  & 150 & 3.8$^{+0.6}_{-0.6}$ & 0.7 $\pm$ 0.4       & LWA; HBA\\
J1645$-$0317          & 0.387690 & 35.76  & 1.32  & 256  & 42.92 -- 66.95  & 327 & 4.1$^{+0.5}_{-0.6}$ & 0.011 $\pm$ 0.004   & LWA; LBA; ORT\\
J1752$-$2806          & 0.562558 & 50.37  & 0.20* & 256  & 66.94 -- 146.48 & 327 & 3.7$^{+0.2}_{-0.2}$ & 1.5 $\pm$ 1.0       & LWA; HBA; ORT\\
J1758+3030            & 0.947256 & 35.07  & 3.28  & 256  & 47.35 -- 76.75  & 370 & 2.2$^{+1.0}_{-1.0}$ & 1.2 $\pm$ 0.5       & LWA; ARECIBO\\
J1823+0550            & 0.752907 & 66.78  & 2.00* & 512  & 47.35 -- 71.85  & 327 & 4.5$^{+1.1}_{-1.4}$ & 0.1 $\pm$ 0.1       & LWA; ORT\\
J1825$-$0935          & 0.769006 & 19.38  & 0.30* & 256  & 42.58 -- 76.76  & 327 & 4.0$^{+1.0}_{-0.7}$ & 0.1 $\pm$ 0.1       & LBA; ORT\\
J1841+0912$\dagger$   & 0.381319 & 49.16  & 1.66  & 128  &       63        & 142 &         --          & 0.3 $\pm$ 0.1       & LBA; HBA \\ 
J1844+1454            & 0.375463 & 41.49  & 1.68  & 256  & 42.13 -- 86.01  & 327 & 4.4$^{+0.3}_{-0.3}$ & 0.09 $\pm$ 0.04     & LWA; ORT\\
J1851+1259            & 1.205303 & 70.63  & 2.64  & 1024 & 111.8 -- 154.8  & 327 & 4.0$^{+0.2}_{-0.2}$ & 1.1 $\pm$ 0.2       & HBA; ORT\\
J1913$-$0440          & 0.825936 & 89.39  & 4.04  & 256  & 66.95 -- 147.75 & 327 & 4.5$^{+0.3}_{-0.5}$ & 4 $\pm$ 3           & LWA; HBA; ORT\\
J1948+3540            & 0.717311 & 129.37 & 7.65  & 1024 & 178.2 -- 610.0  & 925 & 4.5$^{+0.1}_{-0.1}$ & 339 $\pm$ 24        & HBA; ORT; LOVELL\\
J2018+2839            & 0.557953 & 14.20  & 0.98* & 512  & 37.21 -- 57.22  & 327 & 4.4$^{+1.3}_{-1.3}$ & 0.02 $\pm$ 0.01     & LBA; ORT\\
J2055+3630            & 0.221508 & 97.42  & 5.00* & 256  & 145.0 -- 408.0  & 610 & 3.8$^{+0.1}_{-0.2}$ & 45 $\pm$ 5          & HBA; GMRT; ORT; LOVELL\\
J2149+6329            & 0.380140 & 129.72 & 3.88  & 512  & 111.8 -- 170.4  & 186 & 3.0$^{+0.2}_{-0.2}$ & 11 $\pm$ 1          & HBA\\
J2225+6535            & 0.682542 & 36.44  & 1.88  & 256  & 42.41 -- 85.82  & 150 & 4.2$^{+0.6}_{-0.7}$ & 0.1 $\pm$ 0.1       & LWA; HBA\\
J2229+6205            & 0.443055 & 124.64 & 3.01  & 512  & 115.7 -- 170.5  & 610 & 3.2$^{+0.1}_{-0.1}$ & 15 $\pm$ 1          & HBA; LOVELL\\
J2337+6151$\dagger$   & 0.495370 & 58.41  & 0.70  & 512  &       65        & 142 &         --          & 0.2 $\pm$ 0.1       & LBA; HBA \\
\enddata
\end{deluxetable}

\begin{deluxetable}{llcllcc}
\tablecolumns{7}
\tabletypesize{\footnotesize}
\tablecaption{This table lists the pulsars for which the line of sight pass through an 
H{\small II} region or is associated with the supernova remnant. Table lists the pulsar 
name, DM, DM distance (D$_{DM}$), DM independent distance(D$_A$), Association, Distance 
(D$_{H{\small II}}$) and $\alpha$.  The pulsars in the table include $\alpha$ values 
from the literature as given in the caption of Figure~\ref{tdmplot}. The details about 
the pulsar distances are obtained from \citet{psrcat} and the associations are obtained
from H{\small$\alpha$} maps \citep{halpha}.
\label{tab3}}
\tablehead{
PSR J      &      DM      & D$_{DM}$ & D$_A$ & Association & Distance &  $\alpha$ \\
           & pc~cm$^{-3}$ &   (kpc)  & (kpc) &             &    (kpc) &           \\
}
\startdata
J0358+5413 &   57.14   &   1.00      & 1.00$^{+0.20}_{-0.10}$	&	Sh2--205	&	1.00$\pm$0.20	&  4.6$^{+0.7}_{-0.8}$  \\
J0502+4654 &   41.83   &   1.32      &         -- 		&	Sh2--221	&	0.80$\pm$0.40	&  3.5$^{+0.2}_{-0.2}$\\
J0525+1115 &   79.42   &   1.84      &         --		&	Sh2--264	&	0.34		&  3.4$^{+0.4}_{-0.4}$\\
J0534+2200 &   56.77   &   2.00      & 2.00$^{+0.50}_{-0.50}$	&	Crab nebula	&	2.00$\pm$0.50	&  3.4$^{+0.2}_{-0.2}$\\
J0614+2229 &   96.91   &   1.74      &         --		&	Sh2-248		&	1.50		&  2.9$^{+0.1}_{-0.1}$\\
J0742-2822 &   73.78   &   2.00      & 2.00$^{+1.00}_{-0.80}$	&	Gum nebula	&	0.40		&  3.8$^{+0.4}_{-0.4}$\\
J0835-4510 &   67.99   &   0.28      & 0.28$^{+0.02}_{-0.02}$	&	Gum nebula	&	0.40		&  4.5$^{+0.6}_{-0.6}$\\
J0837-4135 &  147.29   &   1.50      & 1.50$^{+1.20}_{-0.90}$	&	Gum nebula	&	0.40		&  2.4$^{+0.1}_{-0.1}$\\
J0840-5332 &  156.50   &   0.57      &         --		&	Gum nebula	&	0.40		&  4.6$^{+0.5}_{-0.5}$\\
J2029+3744 &  190.66   &   5.77      &         --		&	Sh--109		&	1.40		&  4.1$^{+0.2}_{-0.2}$\\
J2055+3630 &   97.42   &   5.00      & 5.00$^{+0.80}_{-0.60}$	&	Sh--109		&	1.40		&  3.8$^{+0.1}_{-0.2}$\\
\enddata
\end{deluxetable}

\end{document}